\documentclass[apjl, iop]{emulateapj}
\usepackage{mathptmx}
\usepackage{setspace}

\slugcomment{{\sc Accepted to ApJL:}  December 23, 2014, {\sc Draft version:} February 10, 2015}

\shorttitle{Direct evidence for an evolving dust cloud from the exoplanet KIC\,12557548\,b}
\shortauthors{Bochinski et al.}

\begin{document}

\title{Direct evidence for an evolving dust cloud from the exoplanet KIC\,12557548\,b}

\author{Jakub J. Bochinski\altaffilmark{1}, Carole A. Haswell\altaffilmark{1}, Tom R. Marsh\altaffilmark{2}, Vikram S. Dhillon\altaffilmark{3} and Stuart P. Littlefair\altaffilmark{3}}

\altaffiltext{1}{Department of Physical Sciences, The Open University,  Milton Keynes, MK7 6AA, UK; jakub.bochinski@open.ac.uk} 

\altaffiltext{2}{Department of Physics, University of Warwick, Coventry, CV4 7AL, UK}

\altaffiltext{3}{Department of Physics and Astronomy, University of Sheffield, Sheffield, S3 7RH, UK}

\begin{abstract}
We present simultaneous multi-color optical photometry using ULTRACAM of the transiting exoplanet KIC\,12557548\,b (also known as KIC\,1255\,b). This reveals, for the first time, the color dependence of the transit depth.
Our $g$' and $z$' transits are similar in shape to the average \emph{Kepler} short-cadence profile, and constitute the highest-quality extant coverage of {\it individual} transits. Our Night 1 transit depths are $0.85\pm 0.04\%$ in $z$'; $1.00 \pm 0.03\%$ in $g$'; and $1.1 \pm 0.3\%$ in $u$'.
We employ a residual-permutation method to assess the impact of correlated noise on the depth difference between the $z$' and $g$' bands and calculate the significance of the color dependence at $3.2 \sigma$.
The Night 1 depths are consistent with dust extinction as observed in the ISM, but require grain sizes comparable to the largest found in the ISM: 0.25--1$\mu$m. This provides direct evidence in favor of this object being a disrupting low-mass rocky planet, feeding a transiting dust cloud.
On the remaining four nights of observations the object was in a rare shallow-transit phase.
If the grain size in the transiting dust cloud changes as the transit depth changes, the extinction efficiency is expected to change in a wavelength- and composition-dependent way. Observing a change in the wavelength-dependent transit depth would offer an unprecedented opportunity to determine the composition of the disintegrating rocky body KIC\,12557548\,b.
We detected four out-of-transit $u$' band events consistent with stellar flares.
\\

\end{abstract}


\section{Introduction}

\citet[][hereafter R12]{rap12} discovered  a peculiar, transit-like signal in the \emph{Kepler} light curve of  KIC\,12557548 (hereafter KIC\,1255), a K5-K7 dwarf star \citep{bor09}. The signal is unlike any exoplanetary transit curve previously observed, characterised by a short pre-ingress brightening phase, a sharp ingress and an extended egress, with a period of just 15.7 hr. The transit depths range from $\lesssim 0.15\%$ to $1.3\%$. R12 proposed that the signal is caused by a highly irradiated super-Mercury, surrounded by an opaque cloud of dust and metal-rich gas constantly replenished by the embedded planet through a thermal Parker-type wind feeding a comet-like obscuring tail. In this scenario, the underlying planet has an undetectable transit signal, and the observed transits are caused by the surrounding cloud of dust only. Alternative explanations, such as a dual giant planet system or an eclipsing binary orbiting KIC\,1255 in a hierarchical triple configuration, were also analysed and ruled out. A similar signal has since been  detected in the light curve of another short period system KOI-2700 \citep{rap14}.

A simple cometary dust model for KIC\,1255\,b was created by \citet[][hereafter B12]{bro12} who estimated an average particle size in the cloud of $\sim0.1\micron$. A pseudo 2D model by \citet[][hereafter W13]{wer13} was applied to individual transits in the  \emph{Kepler} short-cadence data, suggesting that deeper transits tend to have most absorption in the tail, rather than in the disk-shaped, opaque coma surrounding the planet. The pre-transit brightening, attributed to
forward scattering from the dust cloud, was analysed by \citet[][hereafter B13]{bud13}, who estimated the particle radius of the grains in the head of the tail to be about $0.1\micron$ to $1\micron$, likely decreasing in size along the tail.
\citet{kaw13} found a correlation between the transit depth and the activity of the star, pointing to stellar activity being the primary energy source driving the evaporation. Simultaneous \emph{Kepler} and $K_s$-band observations of two transits by \citet[][hereafter C14]{croll14} and time-series optical spectrophotometry with the 10.4\,m GranTeCan (E. Palle 2015, private communication) had insufficient precision to
detect
color dependence in the transit depths at the $\sim 10\%$ level.

In this Letter we discuss high-cadence multi-color observations of KIC\,1255. Our light curves allow us to directly confirm the presence of the dust cloud surrounding the putative planet and establish that with further similar data it could be possible to determine the composition of the scattering material.
We also briefly discuss flaring events observed in the u' band and compare the morphology of individual KIC\,1255\,b transits with the \emph{Kepler} observations.

\section{Observations and Data Reduction}
\label{sec:obs-red}

We observed  KIC\,1255 on 14, 25, 27, 29 and 31 July 2013 (civil, beginning of the night dates) using ULTRACAM \citep{dhi07} on the 4.2m William Herschel Telescope. Dichroic beamsplitters directed incident light into three cameras, each recording a different band. Each light curve was at least 4\,hr long, covering $\sim1.5$\,hr of in-transit and $>2.5$\,hr of out-of-transit time, based on the longest transits of R12. We used Sloan \citep{fuk96} broadband filters: $z$', $g$', $u$' on the first two nights to observe the widest possible wavelength range with high signal-to-noise ratio (S/N) to measure the color dependence of the extinction and scattering.
The transit on the second night was very shallow, approaching the limits of detectability. Consequently we switched from $z$' to $i$' in the red arm of ULTRACAM for the remaining three nights to boost the S/N, while sacrificing some of the wavelength coverage. We used 6.6~s exposure times throughout (co-adding 6 exposures in the $u$' band) to achieve time resolution up to an order of magnitude higher than the \textit{Kepler} short-cadence data (58.9 sec, W13), achieving S/N $> 200$ for $z$', $i$' \& $g$' and $> 50$ for $u$' with duty cycle exceeding 99.6\% due to the use of frame-transfer CCDs in ULTRACAM.
All nights were photometric, with $1''$ seeing on average. Guiding errors, due to flexure in the autoguider relative to ULTRACAM, caused small offsets in R.A. with $\sim0.3''$ (1 pixel) accumulating per hour, which we manually corrected. The flux showed no correlation with the position of the star on the CCDs.

\begin{figure*}
\centering
\epsscale{1.0}
\plottwo{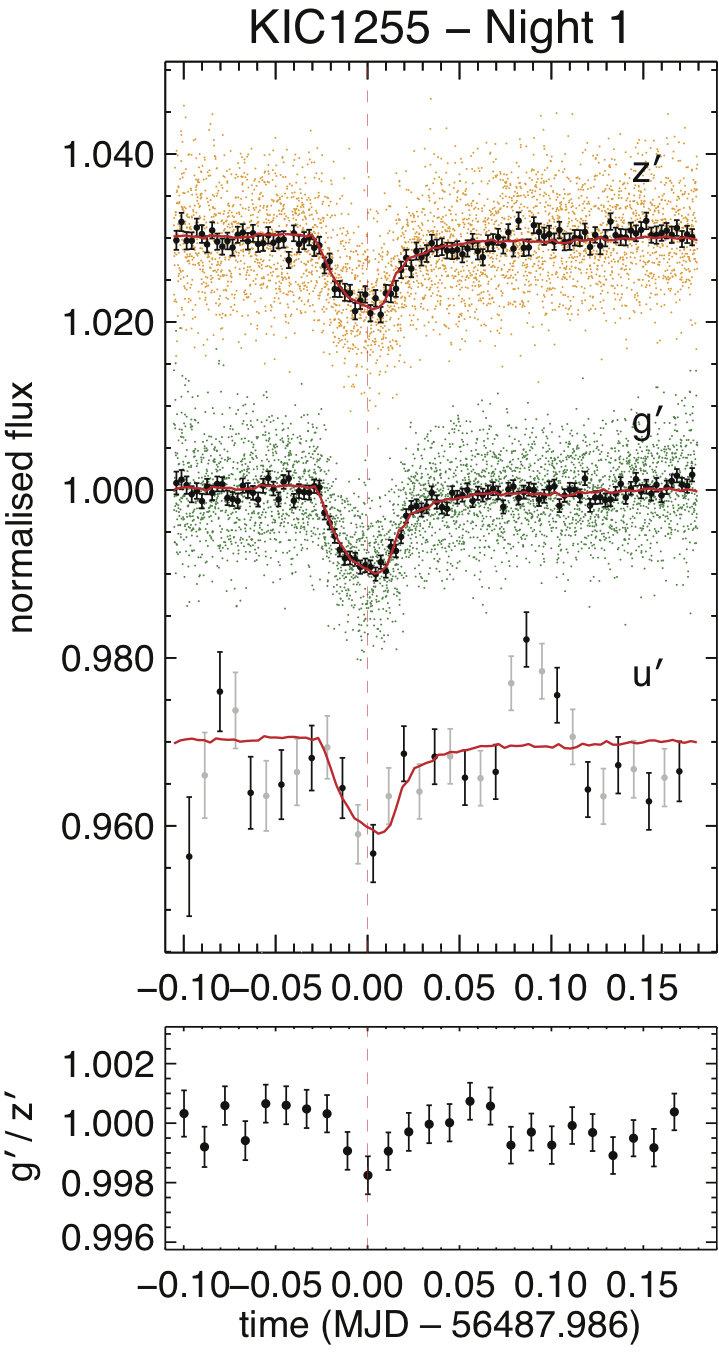}{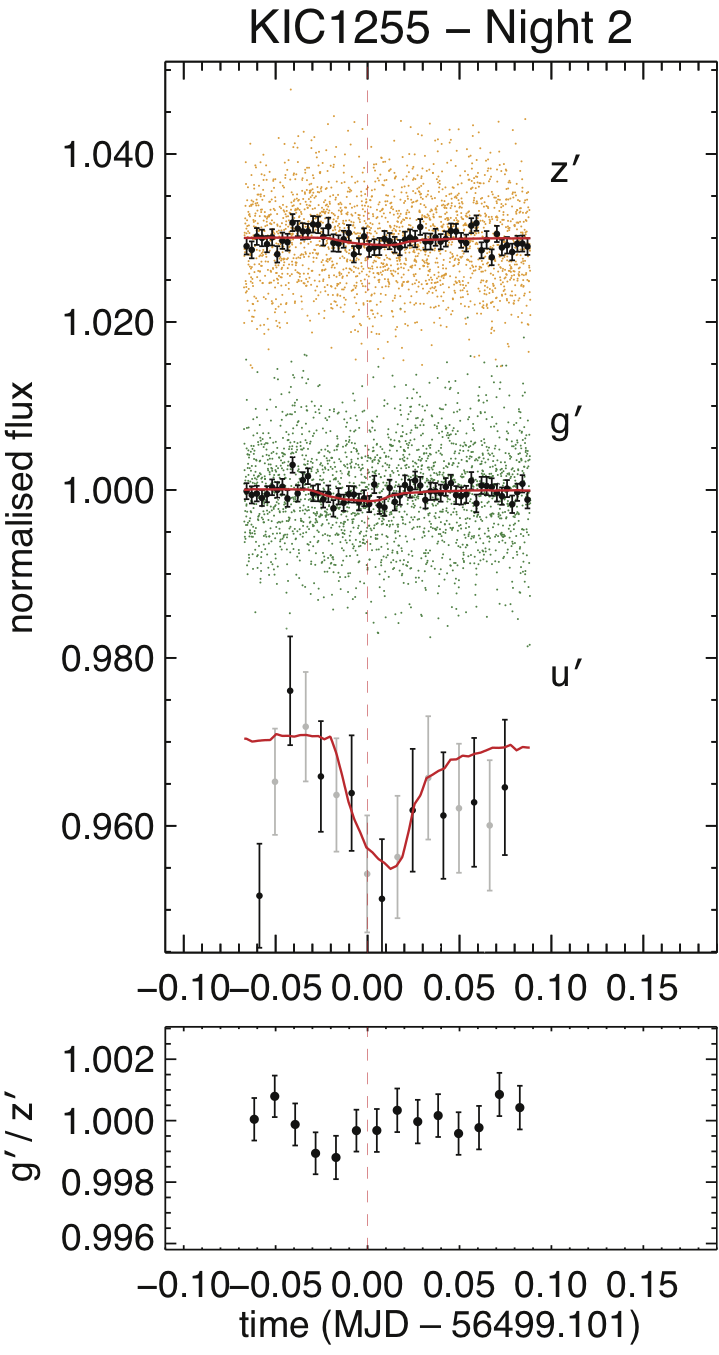}
\caption{\label{lightcurves}Upper panels: $z$' (top), $g$' (centre) and $u$' (bottom) light curves of KIC\,12557548 from 14~\&~25~July~2013; $z$' and $u$' are offset vertically by +0.03 and -0.03 respectively. The scaled \emph{Kepler} short-cadence profile fits are shown in red. four-minute bins ($z$' and $g$') and 24 minute bins ($u$') are shown in black. Grey points ($u$') show
an alternative binning of the same data.
Raw data points, shown in orange for $z$' and green for $g$' are omitted for $u$'. Lower panels: $g$'/$z$' curves.
Times are relative to a zero-point calculated from the fiducial phase of R12 using the period of W13. \\ \\}

\end{figure*}

Frames were calibrated following the standard procedure \citep{how06}, with additional fringing corrections applied to the first two nights of $z$' data. We performed differential aperture photometry using the ULTRACAM pipeline\footnote{http://deneb.astro.warwick.ac.uk/phsaap/software/ultracam/html/\\index.html} and the Improved Differential Photometry method \citep{fer12}, selecting three comparison stars of similar or greater brightness than the target: KIC\,12509691,
KIC\,12509684 and KIC\,12557520. There were no available comparison stars with similar color to KIC\,1255, so we performed a subsequent color term correction by fitting a second-order polynomial, symmetrical with respect to the meridian crossing time, to the out-of-transit section of the light curve. 3-$\sigma$ outliers were omitted during
determination of the color term correction and 8-$\sigma$ outliers were removed from the final results; outliers were fewer than 0.2\% of all data. Figure \ref{lightcurves} shows the resulting light curves for Nights 1 \& 2.

Our light curves show no prominent systematic effects. All $g$'-band light curves for KIC\,1255
are shown in Fig.~\ref{g-4nights}, and similar apparently systematic-free light curves were
obtained in other bands and for our comparison stars. We quantified the correlated noise
in the comparison star light curves and in the out-of-transit  KIC\,1255 data using the time-averaging method \citep{pont06}. The 10--45\,minute range, i.e. the duration of ingress and egress is the most relevant timescale for transit photometry \citep{cw09}; the red noise on this timescale
in our KIC\,1255 Night 1 light curve is on the order of $6\%$, $9\%$, and $30\%$ of the white noise in
$z$', $g$', and $u$' respectively. The noise levels on other nights were comparable, and in the comparison stars
the red noise levels were slightly lower, especially in $u$'. \\

\section{Results and analysis}
\label{data}
\subsection{Transit Characterisation}
Following \citet{has12} we used a simple \citet{man02} planet transit model to determine $R_p/R_*$ and the mid-transit time of KIC\,1255\,b in each band for each night. We set the remaining planetary parameters according to W13 and used stellar parameters from the Kepler Input Catalogue \citep{brw11}. Quadratic limb-darkening coefficients were interpolated from \citet{cla11} using the EXOFAST routine by \citet{eas13}.  We know the transits of KIC\,1255\,b are not attributable to the unchanging circular cross-section of an opaque planet, but this
simple model gives quantitative depth and timing information (see Table~\ref{results}), which can be easily reproduced.

The Mandel \& Agol  fits  to the Night 1 $z$' and $g$' give residuals which show clear systematic deviations, indicating as expected that the dust is distributed differently to a circular cross-section. We also fitted scaled versions of the average \emph{Kepler} short-cadence profile (C14). In this fitting we allowed the time/phase offset to vary freely and applied a scale factor to the flux, so the fractional depth, $d$, at the deepest part of the profile was free to vary, but the duration of the transit profile was fixed.
Like our $g$' and $z$' data, the \emph{Kepler} short-cadence profile has some discontinuities in the slope of the egress, and provides fits which do not exhibit systematic patterns in the residuals.
It is possible that the very smooth egress in the phase-folded
\emph{Kepler} long-cadence curve (R12, compared to our data and to the \emph{Kepler} short-cadence curve) can be partially attributed to the smearing effects of the long \emph{Kepler} exposures.
Our light curves, particularly the $g$' band, which is similar in wavelength coverage to the \emph{Kepler} band, are very similar in shape
to the scaled \emph{Kepler} short-cadence profile.
Figure \ref{lightcurves} shows the
best fitting scaled \emph{Kepler} short-cadence profiles for Nights 1 \& 2,
while Figure \ref{kep-contour} shows the corresponding confidence intervals for Night 1.
Following the rationale in \citet{cw09}, the confidence intervals in Figure~\ref{kep-contour}(a) and Table~\ref{results}
include the contribution from the red noise.

\begin{figure}[t]
\epsscale{1.10}
\plotone{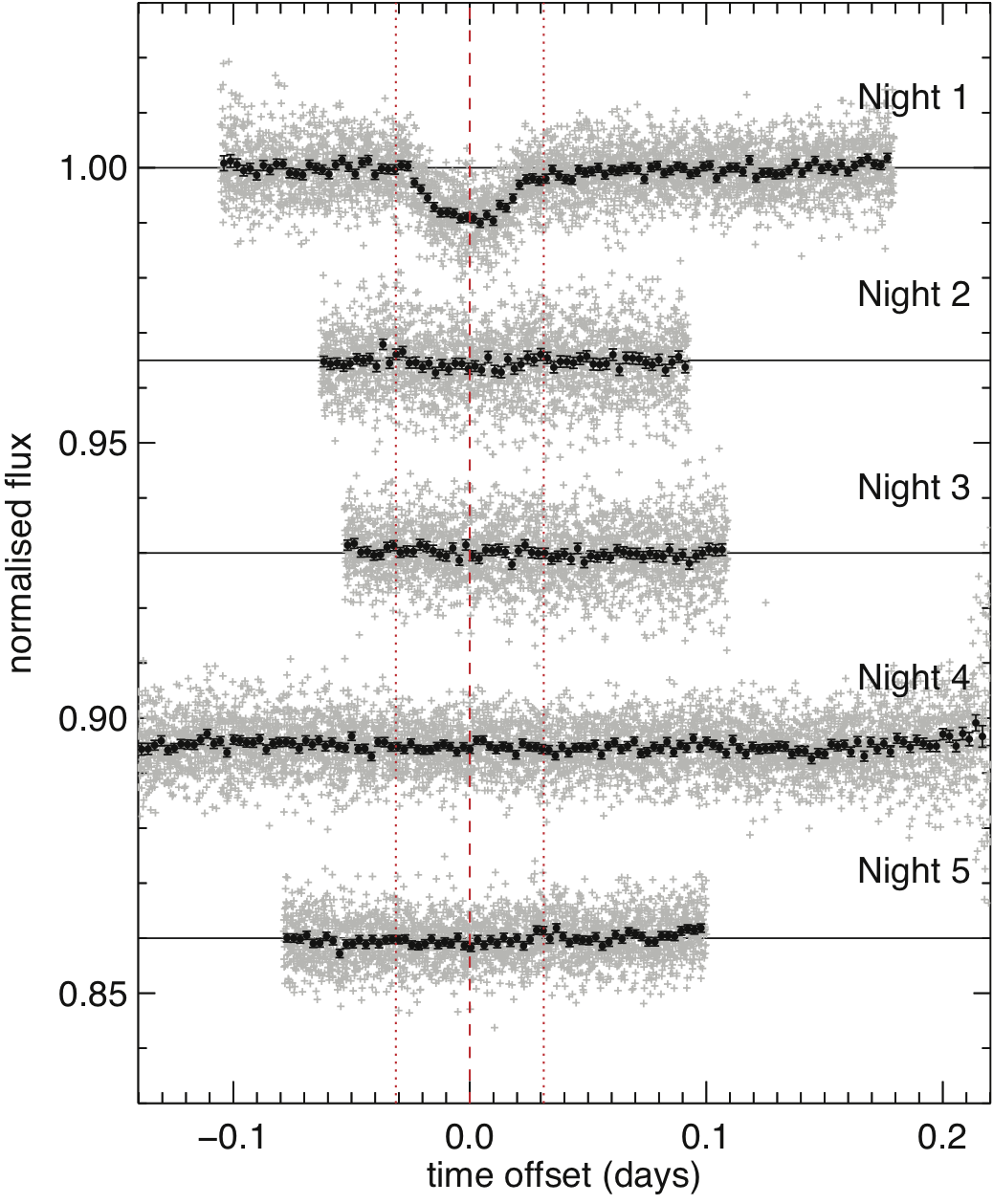}
\caption{\label{g-4nights} Our $g$'-band light curves. Vertical offsets of $(n-1) \times 0.035$, where $n$ is Night number, have been applied. Dashed line: transit time from R12 and W13 (see text). Dotted lines: transit duration (R12). \\}
\end{figure}

\begin{figure*}
\epsscale{1.0}
\plotone{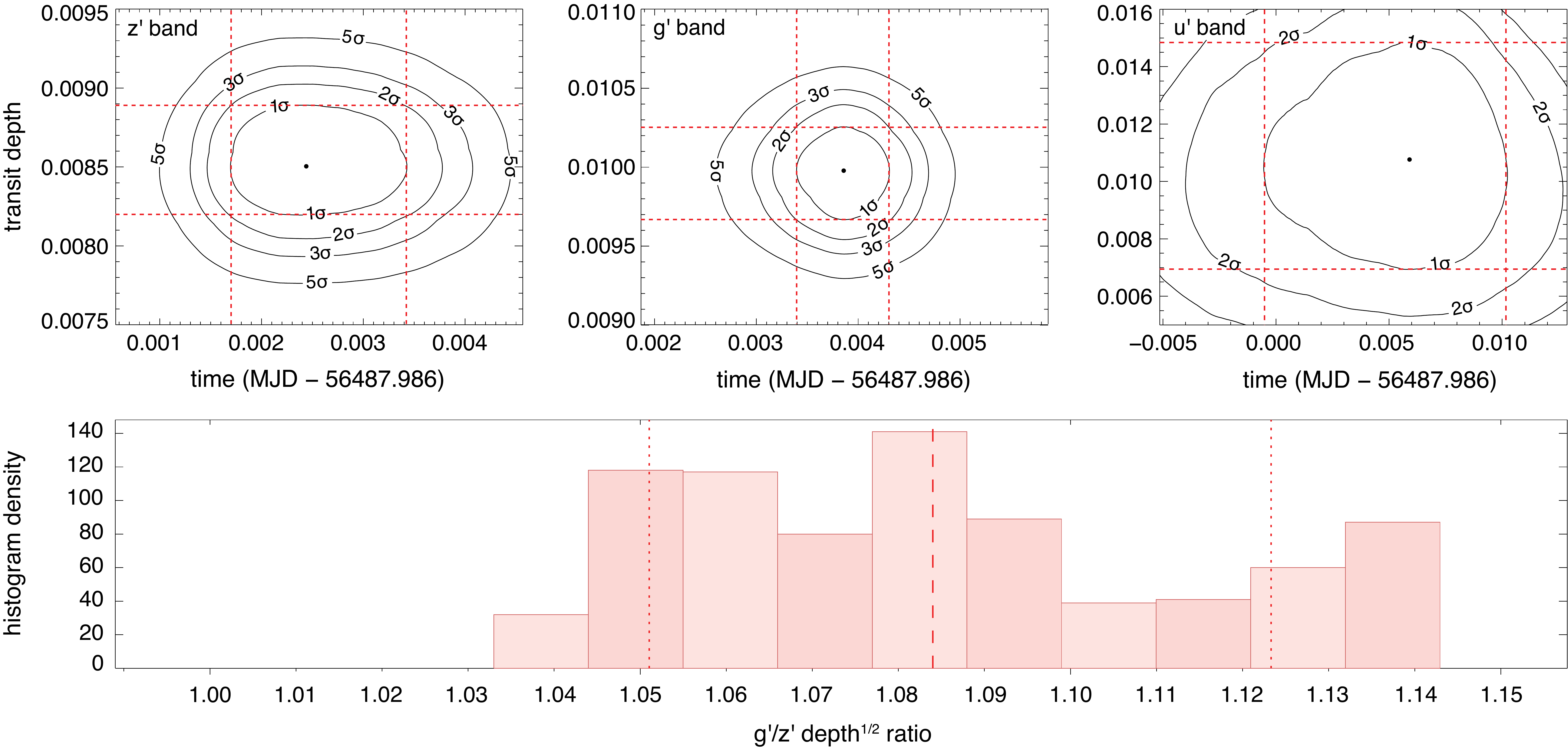}
\caption{\label{kep-contour}(a) Contour maps from our fits of the scaled \emph{Kepler} short-cadence profile to the Night 1 $z$' (left), $g$' (central) and $u$' (right) bands. Black lines give confidence intervals and red dashed lines show 1$\sigma$ uncertainties. (b) Histogram of 807  residual permutations for the $g$'/$z$' depth$^{\frac{1}{2}}$ ratio. The dashed line shows the best fit and the dotted lines show 68.27\% confidence intervals based on the studentised pivotal method \citep{car00}. \\ \\ }
\end{figure*}

We used the (only published) fiducial phase from R12 and the (most precise published) period from W13 to calculate
phase~0.0 independent of our data.
Figure~\ref{lightcurves} shows \emph{time} relative to this.
Where our unbiased search of depth-timing space identifies a transit at the appropriate phase, this is suggestive of a genuine detection. We detected appropriately phased transits on Nights 1 and 2;
the best fit parameters are reported in Table \ref{results}. The Night 2 $g$' light curve (Figures~\ref{lightcurves}~\&~\ref{g-4nights}) clearly has
a noise level similar to the reported transit depth, and the correlated noise can conspire to move the out-of-transit level and the in-transit level in opposite directions, so while the timings suggest this is a genuine detection we caution against placing too much faith in the reported depths.

For Night 1 the reduced $\chi ^2$ values for our Mandel and Agol model fits were
0.872, 0.853, and 0.800 %after cw09
with 4036, 4036 and 671 degrees of freedom
in $z$', $g$' and $u$' respectively;
the corresponding values for the scaled \emph{Kepler} short-cadence profile fits were
0.785, 0.695, and 0.785. % after cw09
The scaled \emph{Kepler} profile obviously provides a superior fit to the $z$' and $g$' data. The $u$' light curve
(Figure~\ref{lightcurves}) clearly does not strongly constrain the shape of the transit.

We report the mid-transit time for the Mandel and Agol fits, while for the \emph{Kepler} profile fits we follow C14 and report $t_{\rm{transit}}$, the time of flux minimum. Because the profile is skewed, $t_{\rm{transit}}$ occurs after the mid-transit time; the Night 1 $z$' and $g$' values in Table~\ref{results} show offsets of ${\approx0.003}$\,days, which is within the uncertainties in the $u$' values.
The depth parameter, $d$, is the fractional depth at the deepest part of the fitted \emph{Kepler} profile. For ease of
comparison with the $R_{\rm p}/{R_*}$ parameter, we tabulate the values of $d^{\frac{1}{2}}$. Again, there is a systematic offset: the broader, flatter floor in the Mandel and Agol transit profile sits above the minimum of the
observed profile, while the \emph{Kepler} profile matches it well. This causes the systematic offset between the
tabulated Night 1 depth parameters of $\approx0.01$ in all three bands.

On Nights 3, 4, \& 5, whose $g$' band light curves are shown in Fiure.~\ref{g-4nights}, we did not recover
appropriately phased transits. We estimated upper limits on  $R_p/R_*$ on these nights
by attempting to recover transits using the method described above, and quoting the largest value of $R_{\rm p}/{R_*}$ identified: 0.031 in $i$', 0.027 in $g$' and  0.087 in $u$' (averaged over three nights). These `transits' were at the wrong phases, indicating the algorithm chose noise features. Any actual transit deeper than these values would have been found if present.

If the obscuring dust is clumpy, we might expect an increased scatter in flux values during ingress and egress as individual clumps
cross the stellar limb. We searched for this effect but found no significant evidence for it. It may, however, emerge if
a larger number of light curves of the quality of our Night 1 data are analysed.

Our primary result (Table \ref{results}) is the color dependence of the transit depth: the ratio of the square roots of the transit depths in $g$'~and~$z$' (i.e. the parameter analogous to $R_{p}/{R_*}$) is $1.084~(+0.024, -0.029)$, where the uncertainties were propagated from the best fit model depths. To assess the robustness of this, we employed residual permutation statistics \citep{win09} with two-sided 68.27\% confidence intervals based on the studentised pivotal method \citep{car00}, concluding that the ratio of the square roots of transit depths in $g$'~and~$z$' is $1.084~ (+0.039, -0.033)$. As none of the 807 realisations of the residual permutation method produced a depth ratio of 1 (Figure \ref{kep-contour}(b)), we reject the zero color difference hypothesis with probability 99.88\% or $3.2\sigma$.

To assess whether the transit depth, which can be influenced by limb darkening, is a good proxy for light loss due to extinction in a given band, we compared the $g$'/$z$' depth ratio described above to the ratio of the equivalent occulting areas $A_{g'} / A_{z'}$ in these two bands. 
To find $A_{g'}$~\&~$A_{z'}$, we modelled the occulting dust tail as a cloud with variable transparency extending
$0.075 R_{*}$ above and below the orbital plane and $60^{\circ}$ following the orbit of the planet (B13). The dust density profile of the tail, inversely proportional to its transparency, is: 
\begin{equation}
\rho(\theta) = \rho(0) e^{C\theta / \pi}
\end{equation}
i.e. Equation 4 from B13; where  $\theta$ is the angle from the beginning of the arc, $\rho(\theta)$ is the density along the arc and $C$ is a constant. $C = -25$ produces a good match to the transit light curve in all bands, and is consistent with B13. The transparency along the tail was adjusted by varying $\rho(0)$ in each band separately, until the light curves matched the observations. The square root of the resulting $A_{g'} / A_{z'}$ ratio is $1.077 \pm 0.001$ (the uncertainty arises from a reasonable range of limb-darkening coefficients), well within the confidence intervals for the square root depth ratio. This establishes that the impact of limb darkening on our results is minor in comparison to other sources of error.

The presence of occulted and unocculted starspots can modify the transit depth and possibly influence the derived dependence of the planetary radius on the wavelength \citep{pon08}. To assess this, we looked at analogous multi-wavelength observations of a well studied mid-K star, HAT-P-20 \citep{bak11}, and used Figure~4 in \citet{bal12} to calculate the effect of starspots on the theoretical HAT-P-20 $g$'/$z$' transit depth ratio. The maximum fractional light loss due to starspots was set at 4\% as predicted by \citet{croll14b} for KIC\,1255. We found that starspots can account for only up to 0.46\% of our observed $g$'/$z$' square root depth ratio: insignificant in comparison to other sources of error.

%\subsection{The updated Ephemeris}

%We find a new ephemeris for KIC\,1255\,b based on two epochs: 55399.824(2)~MJD derived from R12 and 56487.9861(5)~MJD from our Night 1 z' band results (Table \ref{results}). The resulting new period 0.653551(2)d agrees well with estimates by R12 of 0.65356(1)d and B13 of 0.6535521(15)d, while being slightly shorter than the more recent estimate by \citet{wer13} of 0.6535538(1)d. Based on our estimates, for the purpose of calculating future transit times:
%\begin{equation}
%t_n = 56487.9861(5) + n \times 0.653551(2) \mbox{ MJD}
%\end{equation}

%To search for a potential change in period over time, we compared our period estimate with those by R12, W13 and B13 but detected only a slight negative trend $\dot{P} = - 0.3 \pm 0.4 $ sec/year clearly consistent with no period change at all. Further monitoring of the transits, to see if the trend exists, is needed.

\subsection{Color Dependence and Dust Modelling}
\label{dust}

We observed differences in transit depths between the $u$', $g$' and $z$' filters on Night 1,
with an increase in depth toward shorter wavelengths
(Figures~\ref{lightcurves}~\&~\ref{kep-contour}, Table \ref{results}). This effect is consistent with extinction from the putative dust cloud surrounding the planet (R12), as we will show quantitatively. Detections on Night 2 (Figure~\ref{lightcurves}, Table \ref{results})
also show an increase in depth towards shorter wavelengths.

\begin{figure}
\epsscale{1.15}
\plotone{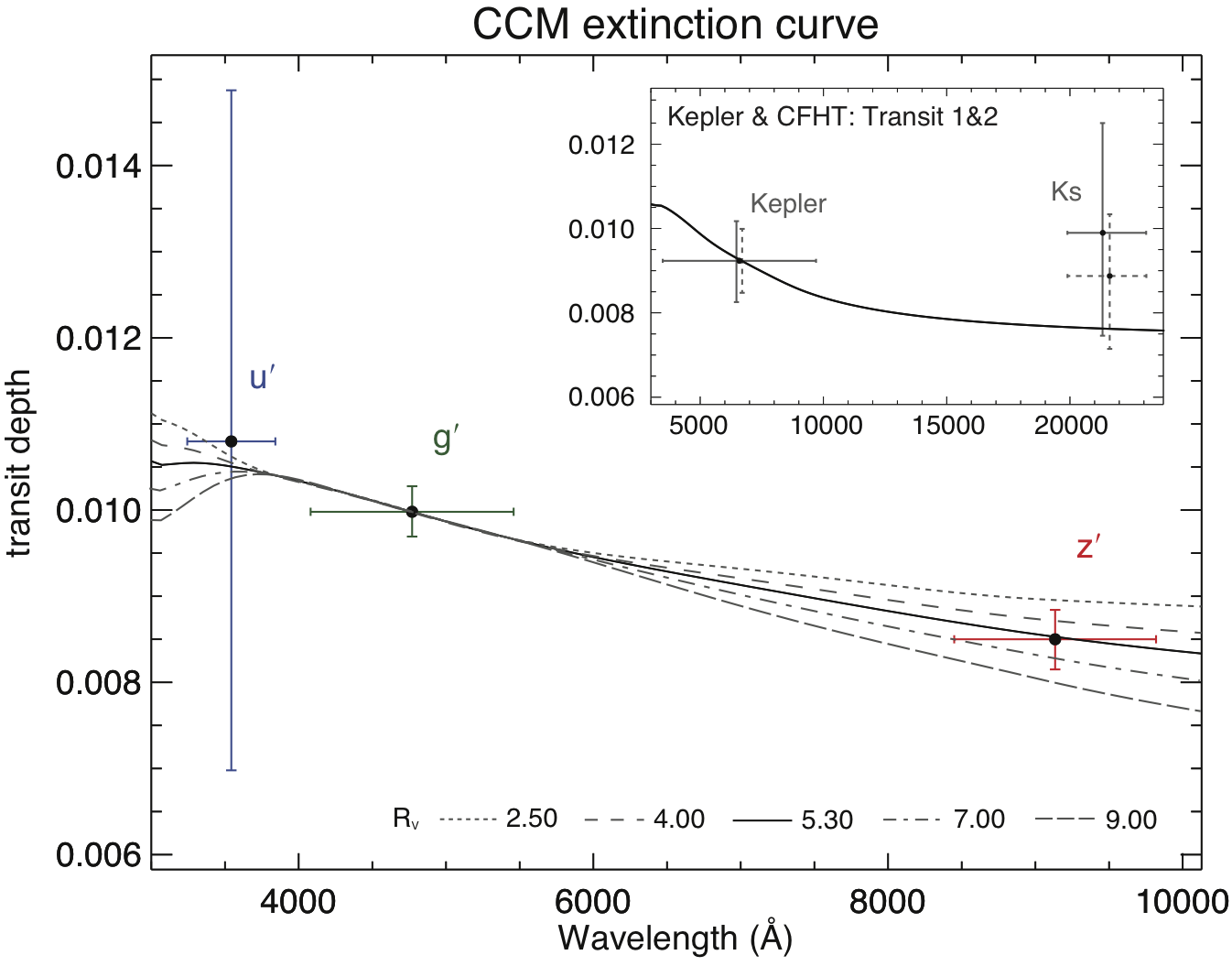}
\caption{\label{ccm} The CCM89 ISM extinction law fitted to our Night 1 transit depths. The fit included two free parameters: E(B-V) and R$_V$. The best fit values were E(B-V)=0.056 (used throughout) and R$_V= 5.3$ (solid line). Extinction laws for values of R$_V$ ranging between 2.50 and 9.00 are also shown. The inset shows our best fit CCM law and C14's photometry normalized to the same transit depth; it is consistent with our fit. C14's Night 2 vertical error bars (dashed lines) have been offset slightly in wavelength for visibility. }
\end{figure}

We modelled our Night 1 scaled \emph{Kepler} depths
with the \citet[][hereafter CCM89]{ccm89} interstellar extinction law to find that the extinction depths are broadly
consistent with the wavelength-dependence of the reddening seen in the ISM (Fig. \ref{ccm}). This is strong independent and direct evidence in favor of the dust-cloud model for the transits. We constrained $R_V$ to the highest values found in the ISM: $R_V \approx 5.3$. As $R_V$ increases with the mean dust grain size \citep{mw81}, the grain sizes we found in the KIC\,1255 dust cloud on Night 1
are comparable to the largest found in the ISM \citep[0.25--1$\micron$ depending on the grain composition,][]{mat77}. We normalised and plotted the two transits from C14 in the inset to Figure~\ref{ccm}; their color dependence is consistent with our fit, as well as with a constant transit depth. Note that the C14 transits were observed on a different night to our observations and hence a different grain size distribution might apply. The transits vary because the dust cloud is varying, and it is difficult to imagine a scenario where the irradiated dust cloud changes on planet-sized scales while the dust grains comprising it retain an identical size distribution.

\begin{table}[b]
\caption{\label{results}Parameters derived from fitting the Mandel \& Agol model and the averaged \emph{Kepler} profile.}
\footnotesize
{\renewcommand{\arraystretch}{1.5}
\begin{tabular}{l  c  c  c  c}
\hline \hline  
Filters & \multicolumn{2}{ c }{Night 1} & \multicolumn{2}{ c }{Night 2} \\  
\hline \\ [-2.5ex]
& $R_p/R_*$ & $d^{\frac{1}{2}}$ \textsuperscript{{\footnotesize $\dagger$}} & $R_p/R_*$ & $d^{\frac{1}{2}}$  \textsuperscript{{\footnotesize $\dagger$}}\\ 
~~~$z$' & 0.0826 $^{+0.0013}_{-0.0016}$ & 0.0922 $^{+0.0020}_{-0.0016}$ & 0.023 $^{+0.006}_{-0.006}$ & 0.029 $^{+0.007}_{-0.007}$ \\
~~~$g$' & 0.0891 $^{+0.0014}_{-0.0010}$ & 0.0999 $^{+0.0013}_{-0.0015}$ & 0.035 $^{+0.004}_{-0.004}$ &  0.037 $^{+0.006}_{-0.006}$\\
~~~$u$' & 0.090 $^{+0.016}_{-0.020}$ & 0.104 $^{+0.018}_{-0.020}$ & 0.097 $^{+0.031}_{-0.043}$ & 0.137 $^{+0.033}_{-0.031}$ \\ \\ [-2.5ex]
\hline \\ [-2.8ex]
~$g$'/$z$' & 1.079 $^{+0.027}_{-0.021}$ & ~~1.084 $^{+0.024}_{-0.029}$ \textsuperscript{{\footnotesize *}}  & 1.52 $^{+0.43}_{-0.43}$ & 1.28 $^{+0.37}_{-0.37}$\\ \\ [-2.8ex]
\hline \\ [-2.5ex]
& mid-transit & $t_{\rm{transit}}$\textsuperscript{{\footnotesize $\ddagger$}}& mid-transit & $t_{\rm{transit}}$\textsuperscript{{\footnotesize $\ddagger$}}\\
~~~$z$' & 56487.9858  & 56487.9886  & 56499.103  & 56499.110 \\ [-1ex]
 & $^{+0.0004}_{-0.0005}$ & $^{+0.0010}_{-0.0007}$ & $^{+0.010}_{-0.001}$ & $^{+0.006}_{-0.005}$ \\
~~~$g$' & 56487.9866 & 56487.9901  & 56499.092  & 56499.101\\ [-1ex]
 &  $^{+0.0004}_{-0.0004}$ & $^{+0.0004}_{-0.0004}$ & $^{+0.006}_{-0.001}$ & $^{+0.005}_{-0.002}$\\
~~~$u$' & 56487.986  & 56487.992  & 56499.106  & 56499.113 \\ [-1ex]
 & $^{+0.007}_{-0.004}$ & $^{+0.004}_{-0.007}$ & $^{+0.007}_{-0.004}$ & $^{+0.005}_{-0.005}$\\ \\ [-2.5ex]
\hline \hline \\ [-2.5ex]

\end{tabular}}

%~~~z' & 56487.9858 $^{+0.0004}_{-0.0005}$ & 56487.9886 $^{+0.0010}_{-0.0007}$ & 56499.103 $^{+0.010}_{-0.001}$ & 56499.110 $^{+0.006}_{-0.005}$\\
%~~~g' & 56487.9866 $^{+0.0004}_{-0.0004}$ & 56487.9901 $^{+0.0004}_{-0.0004}$ & 56499.092 $^{+0.006}_{-0.001}$ & 56499.101 $^{+0.005}_{-0.002}$\\
%~~~u' & 56487.986 $^{+0.007}_{-0.004}$ & 56487.992 $^{+0.004}_{-0.007}$ & 56499.106 $^{+0.007}_{-0.004}$ & 56499.113 $^{+0.005}_{-0.005}$\\

{\footnotesize \textsuperscript{{\footnotesize $\dagger$}} Square root of the transit depth from \emph{Kepler} averaged profile fitting. Comparable to the $R_p/R_*$ value.} \\
{\footnotesize \textsuperscript{{\footnotesize $\ddagger$}} Transit time as defined in \citet{croll14}}\\
{\footnotesize \textsuperscript{{\footnotesize *}} Alternative result derived from the residual-permutation method is 1.084~$^{+0.039}_{-0.033}$      }
\end{table}

\subsection{Flares}
\label{flares}
We observed two potential flaring events in the $u$' band on Night 1
(Figure~\ref{lightcurves}), characterised by a sudden increase in brightness by approximately $1.0\%$ and $1.5\%$
with a similarly abrupt drop back to base level.
On each of Nights 2 and 4 there were single similar events with amplitudes $1.0\%$ and $1.5\%$ respectively.
None of our comparison stars showed anything similar, suggesting these are genuine
astrophysical signals originating from KIC\,1255. The atypical flare shape
can be attributed
to the faintness of the target coupled with the low effective cadence of our $u$' band data hiding the
exponential decay. On Night 1, the second flare overlaps in phase with a slight brightening in $z$' band. Similarly on Night 2, a bright flare in the $u$' band overlaps in phase with a brightening in both the $z$' and $g$' bands. To quantify the amount of extra emission expected in the $u$', $g$' and $z$' bands, we modelled continuum or `white light' flare emission as a blackbody curve with a temperature of 10,000\,K and a variable fill factor, defined as the effective surface area of the star covered by the active region \citep{kow10}. We added this to the model spectral energy distribution for KIC\,1255 \citep{how11} and convolved the resulting artificial spectrum with filter responses for the \emph{SDSS} and \emph{Kepler} bands. The best fit to the observed $\sim1\%$ $u$' band flare signals was achieved for a fill factor of 0.013\%, with a surplus flare to stellar flux ratio in the $u$', $g$', $z$' and \emph{Kepler} bands equal to $1.0\%$, $0.3\%$, $0.1\%$ and $0.5\%$ respectively. This shows that flares originating on KIC\,1255 are expected to be brightest in the $u$' band, with signal in the remaining \emph{SDSS} bands comparable to our noise level.

The calculated surplus emission in the \emph{Kepler}  band is consistent with `white light' flares observed on other K~dwarfs \citep[between $0.1\%$ and $1.0\%$;][]{wal11}. If flares on KIC\,1255 are similarly `white' in nature, they should be borderline visible in the \emph{Kepler} short-cadence photometry.  We inspected all currently accessible short-cadence Kepler photometry (Q13-Q17) and found several similar brightening effects with an amplitude of up to 0.5\% and duration between 30 and 90 minutes. These are consistent with our predictions, but also comparable to the \emph{Kepler} photometric precision at $\sim0.4\%$. Additionally, a 1200 days long cadence Kepler light curve of KIC\,1255\,b shows at most three significant outliers that could be consistent with flaring events or simply outliers, the strongest having an amplitude of 3.5\% (S. Rappaport \& B. Croll 2015, private communication). \citet{kaw13} found that KIC\,1255 is capable of producing even brighter flares, with a brightness increase of up to $\sim6\%$ on timescales of 30 minutes.

\section{Discussion}
We obtained multi-color optical light curves of the transits of KIC\,1255\,b. For the first time these transits are measured with sufficient precision to discern features of
amplitude $0.2\%$ and duration $\sim 4$~minutes
in an individual transit light curve. This permitted
robust quantitative characterisation of the transit shape and depth on Night 1.
We clearly detect a difference in depth between the $z$' and $g$' bands: visible in
the lower panel of Figure~\ref{lightcurves} and quantified in Table~1 and Figure~\ref{kep-contour}. We rule out the effects of limb darkening and starspots as significant contributions to this depth difference.
The color dependence of the transit depth is consistent with extinction due to dust with grain sizes $0.25 \mu{\rm m}  < a <  1\mu \rm{m}$, where $a$ is the grain radius.

Eleven nights later, on Night 2  we detected much shallower $z$' and $g$' transits, with no detection of color dependence in the $g$' and $z$' depths (Table~\ref{results}).
The changing transit depth is probably primarily caused by changes in the quantity and spatial distribution of
dust due to a limit-cycle behavior of the dust injection mechanism (R12).
Clearly the dust cloud produces less extinction during the shallow and undetected transits than it did on Night 1.
It would be very interesting to examine the color dependence of the transits at a variety of depths: this will reveal whether the grain size distribution changes with transit depth.

Figure 13 of C14 shows dust grain extinction efficiency as a function
of $X = 2 \pi a / \lambda$ for grain size $a$ and wavelength $\lambda$. For  $X > \sim 2$ the efficiency plateaus at its peak value irrespective of composition.
For smaller values of $X$ the extinction efficiency drops off in a composition-dependent way. As $X \propto \lambda^{-1}$, this lessening of extinction efficiency with decreasing grain size would be
most pronounced at long wavelengths.
With further observations of quality comparable to Night 1, with transits of varying depths but deep enough for secure detections, we have the prospect of identifying the composition of the dust grains
by matching the observed wavelength-dependent depth changes to one of the curves in Figure 13 of C14.
In our 2013 campaign, we were extremely unlucky to observe rare shallow transits with depth $ < 0.07\%$  (in $g$') on four of the five nights we observed. Only two such quiescent periods were observed by W13 in years 2009 - 2013, amounting to $\sim3\%$ of total \emph{Kepler} observing time.

We have presented direct evidence that the variable transits in KIC\,1255\,b are due to an evolving cloud of dust attributable to mass loss from
a disintegrating low-mass rocky planet.

Analysis of a larger number of transit light curves potentially offers the opportunity to
measure the composition of the grains. This is an unprecedented opportunity to learn the make-up of a rocky extrasolar planet: further observations are urgently required.

\acknowledgments
We are grateful to the referees Saul Rappaport and Bryce Croll for comments and suggestions which substantially improved this paper,
and for providing the average \emph{Kepler} short-cadence profile used in our fitting. We also thank J. Birkby for providing calibration data and D. Staab for useful comments. J.J.B. is supported by an STFC studentship, C.A.H., T.R.M., V.S.D., S.P.L. and ULTRACAM are supported by STFC under grants ST/L000776/1, ST/L000733/1, ST/J001589.

{\it Facilities:} \facility{ING:Herschel}.\\ \\ \\


\begin{thebibliography}{}
\bibitem[Ballerini et al.(2012)]{bal12} Ballerini, P., Micela, G., Lanza, A. F., \& Pagano, I. 2012, \aap, 539, A140
\bibitem[Bakos et al.(2011)]{bak11} Bakos, G. A., Hartman, J., Torres, G., et al. 2011, \apj, 742, 116
\bibitem[Borucki et al.(2009)]{bor09} Borucki, W. J., Koch, D., Jenkins, J., et al. 2009, Science, 325, 709
\bibitem[Brogi et al.(2012)]{bro12} Brogi, M., Keller, C. U., de Juan Ovelar, M., et al. 2012, \aap, 545, L4 (B12)
\bibitem[Brown et al.(2011)]{brw11} Brown, T. M., Latham, D. W., Everett, M. E. \& Esquerdo, G. A. 2011, \aj, 142, 11
\bibitem[Budaj(2013)]{bud13} Budaj, J.  2013, \aap, 557, A72 (B13)
\bibitem[Carpenter \& Bithell(2000)]{car00} Carpenter, J. \& Bithell, J. 2000, Statist. Med., 19, 1141-1164
\bibitem[Carter \& Winn(2009)]{cw09} Carter, J. A. \& Winn, J. N. 2009, \apj, 704, 51
\bibitem[Cardelli, Clayton \& Mathis(1989)]{ccm89} Cardelli, J. A., Clayton, G. C. \& Mathis, J. S. 1989, \aj, 345, 245 (CCM89)
\bibitem[Claret \& Bloemen(2011)]{cla11} Claret, A. \& Bloemen, S. 2011, \aap, 529, A75
\bibitem[Croll et al.(2014a)]{croll14} Croll, B., Rappaport, S., DeVore, J., et al. 2014a, \apj, 786, 100 (C14)
\bibitem[Croll et al.(2014b)]{croll14b} Croll, B., Rappaport, S. \& Levine, A. M. 2014b, arXiv:1410.4289
\bibitem[Dhillon et al.(2007)]{dhi07} Dhillon, V. S., Marsh, T. R., Stevenson, M. J., et al.  2007, \mnras, 378, 3
\bibitem[Eastman et al.(2013)]{eas13} Eastman, J., Gaudi, B. S. \& Agol, E. 2013, \pasp, 125, 83
\bibitem[Fern\'{a}ndez Fern\'{a}ndez et al.(2012)]{fer12} Fern\'{a}ndez Fern\'{a}ndez, J., Chou, D.-Y., Pan, Y.-C., \& Wang, L.-H. 2012, \pasp, 124, 915
\bibitem[Fukugita et al.(1996)]{fuk96} Fukugita, M., Ichikawa, T., Gunn, J. E., et al. 1996, \aj, 111, 1748
\bibitem[Haswell et al.(2012)]{has12} Haswell, C. A., Fossati, L., Ayres, T., et al. 2012, \aj, 760, 79
\bibitem[Howarth(2011)]{how11} Howarth, I. D. 2011, \mnras, 413, 3
\bibitem[Howell(2006)]{how06} Howell, S. B. 2006, Handbook of CCD astronomy (2nd ed.; Cambridge, UK: CUP)
\bibitem[Kawahara et al.(2013)]{kaw13} Kawahara, H., Hirano, T., Kurosaki, K., Ito, Y. \& Ikoma, M. 2013, \apjl, 776, L6
\bibitem[Kipping(2010)]{kip10} Kipping, D. M., 2010, \mnras, 408, 3
\bibitem[Kowalski et al.(2010)]{kow10} Kowalski, A. F., Hawley, S. L., Holtzman, J. A., Wisniewski, J. P. \& Hilton, E. J. 2010, \apjl, 714, 1
\bibitem[Mandel \& Agol(2002)]{man02} Mandel, K. \& Agol, E.  2002, \apjl, 580, L171
\bibitem[Mathis et al.(1977)]{mat77} Mathis, J. S., Rumpl, W. \& Nordsieck, K. H. 1977, \apj, 217, 425
\bibitem[Mathis \& Wallenhorst(1981)]{mw81} Mathis, J. S. \& Wallenhorst, S. G. 1981, \apj, 244, 483
\bibitem[Perez-Becker \& Chiang(2013)]{per13} Perez-Becker, D. \& Chiang, E. 2013, \mnras, 433, 3
\bibitem[Pont, Zucker \& Queloz(2006)]{pont06} Pont, F., Zucker, S. \& Queloz, D. 2006, \mnras, 373, 231
\bibitem[Pont et al.(2008)]{pon08} Pont, F., Knutson, H., Gilliland, R. L., Moutou, C. \& Charbonneau, D. 2008, \mnras, 385, 109
\bibitem[Rappaport et al.(2012)]{rap12} Rappaport, S., Levine, A., Chiang, E., et al. 2012, \apj, 762, 1 (R12)
\bibitem[Rappaport et al.(2014)]{rap14} Rappaport, S., Barclay, T., DeVore, J., et al. 2014, \apj, 784, 1
\bibitem[Walkowicz et al.(2011)]{wal11} Walkowicz, L. M., Basri, G., Batalha, N., et al. 2011, \apj, 141, 2
\bibitem[van Werkhoven et al.(2013)]{wer13} van Werkhoven, T., Brogi, M., Snellen, I. A. G. \& Keller, C. U.  2013, \aap, 561, A3 (W13)
\bibitem[Winn et al.(2009)]{win09} Winn, J. N., Holman, M. J., Henry, G. W., et al. 2009, \apj, 693, 794

\end{thebibliography}
\end{document}